\newcommand{\FigPath}[1]{./#1}
\newcommand{\short}[1]{\textsl{#1}}

\documentclass[prl,twocolumn,amsmath,amssymb,showpacs,noshowkeys,superscriptaddress]{revtex4}

\usepackage{graphicx}
\usepackage{dcolumn}
\usepackage{bm}

\begin{document}

\title{Equilibrium First-Order Melting and Second-Order Glass Transitions of the Vortex Matter in
Bi$_2$Sr$_2$CaCu$_2$O$_8$}

\author{H.~Beidenkopf}
\email{haim.beidenkopf@weizmann.ac.il}
\author{N.~Avraham}
\author{Y.~Myasoedov}
\author{H.~Shtrikman}
\author{E.~Zeldov}
\affiliation{Department of Condensed Matter Physics, Weizmann
Institute of Science, Rehovot 76100, Israel}
\author{\\B.~Rosenstein}
\affiliation{Department of Condensed Matter Physics, Weizmann Institute of Science, Rehovot 76100, Israel}%
\affiliation{National Center for Theoretical Sciences and Electrophysics Department, National Chiao Tung University, Hsinchu 30050, Taiwan, R.O.C.}%
\author{E.H.~Brandt}
\affiliation{Max-Planck-Institut f\"{u}r Metallforschung,
Heisenbergstr. 3, D-70506 Stuttgart, Germany}
\author{T.~Tamegai}
\affiliation{Department of Applied Physics, The University of Tokyo, Hongo, Bunkyo-ku, Tokyo 113-8656, Japan}%
\date{\today}

\begin{abstract}
The thermodynamic $H-T$ phase diagram of Bi$_2$Sr$_2$CaCu$_2$O$_8$
was mapped by measuring local \emph{equilibrium} magnetization
$M(H,T)$ in presence of vortex `shaking'. Two equally sharp
first-order magnetization steps are revealed in a single
temperature sweep, manifesting a liquid-solid-liquid sequence. In
addition, a second-order glass transition line is revealed by a
sharp break in the equilibrium $M(T)$ slope. The first- and
second-order lines intersect at intermediate temperatures,
suggesting the existence of four phases: Bragg glass and vortex
crystal at low fields, glass and liquid at higher fields.
\end{abstract}

\pacs{74.25.Qt, 74.25.Dw, 74.72.Hs, 64.70.Pf}


\keywords{}

\maketitle


The magnetic field vs.~temperature ($H-T$) phase diagram of the
vortex matter in high-temperature superconductors, and in
Bi$_2$Sr$_2$CaCu$_2$O$_8$ (BSCCO) \short{in particular}, has drawn
extensive scientific attention \cite{blatter1}. The commonly cited
thermodynamic phase diagram of BSCCO currently consists of a
single unified first-order (FO) melting line. It separates the
low-field quasi long-range ordered Bragg glass (BrG) phase from
the high-field liquid and glass phases \cite{blatter1, avraham1,
radzyner1, vinokur1, mikitik1, giamarchi2}. It is not clear,
however, whether the two high-field disordered phases are
thermodynamically distinct, or rather reflect a gradual dynamic
cross-over \short{from liquid into a frozen, pinned state upon
cooling} \cite{reichhardt1, nonomura1}. In this letter we show
that the equilibrium phase diagram of the vortex matter is indeed
more diverse than the one usually considered.

Experimentally, one of the main obstacles in mapping the
low-temperature thermodynamics of the vortex matter is its
logarithmically slow relaxation rate. Consequently, the phase
diagram has been studied in the past mostly through dynamic
phenomena. \short{Two} prime examples are the irreversibility line
itself, \short{marking} the onset of hysteresis, and the second
magnetization peak (SMP), observed along such hysteretic
magnetization loops \cite{khaykovich2}.

Recently, though, vortex `shaking' was shown to be extremely
effective in catalyzing relaxation at low temperatures
\cite{avraham1, willemin1, brandt1}. Its application unveiled the
inverse melting and the thermodynamic FO transition as the
phenomenon underlying the non-equilibrium SMP \cite{avraham1}. The
`shaking' method employs the segregated penetration of an in-plane
field component into the highly anisotropic BSCCO samples in the
form of Josephson vortices, which are confined in between the
CuO$_2$ planes. In the presence of an \emph{ac} in-plane field,
the Josephson vortices instantaneously bisect the pancake vortex
(PV) stacks on their passage, interacting mainly with adjacent
PVs, while most of the PVs in the stack remain at rest
\cite{koshelev1}. These occasional interactions agitate pinned
PVs, assisting them in assuming their equilibrium configuration.

Within the present study we performed local magnetization
measurements by field and temperature sweeps, while utilizing the
`shaking' method, to map the \emph{equilibrium} phase diagram of
the vortex matter. \short{The} cross-mapping of the FO melting
line along both sweeping directions \short{shows an} excellent
agreement. Temperature sweeps provided particularly sharp
features, with which we demonstrate a liquid-solid-liquid sequence
of phases. Erta\c{s} and Nelson \cite{nelson1} have predicted such
a liquid-solid-liquid sequence to occur within a single
temperature sweep, but it was never observed experimentally. We
further find evidence of a \short{novel second order (SO) phase
transition} within the vortex solid phase, which bears important
consequences regarding the nature of the BrG phase.

The reported results were obtained with a slightly over-doped
BSCCO crystal with $T_\text{c} \approx$ \mbox{90 K} grown by the
travelling solvent floating zone method \cite{motohira1}. This
specific sample was polished into a triangular prism of base
\mbox{$660 \times 270 \: \mu \text{m}^2$} and height \mbox{$70 \:
\mu \text{m}$} \cite{majer1} (other samples yielded similar
results, to be presented elsewhere). The sample was attached onto
an array of eleven \mbox{$10 \times 10 \: \mu \text{m}^2$}
GaAs/AlGaAs Hall sensors. In all measurements taken below \mbox{60
K} the sample was subject to a \mbox{10 Hz} in-plane \textsl{ac}
field of amplitude \mbox{350 Oe}, which was aligned parallel to
the planes to an accuracy  of a few millidegrees. Note that
according to the anisotropic scaling theory \cite{blatter1} this
in-plane field is effectively attenuated by a factor $\gamma
\simeq 200$ - the anisotropy constant in BSCCO. We found that at
higher temperatures `shaking' had no effect on the FO transition
besides a small broadening (see below).

\begin{figure} [!b]
\includegraphics[width=0.38\textwidth]{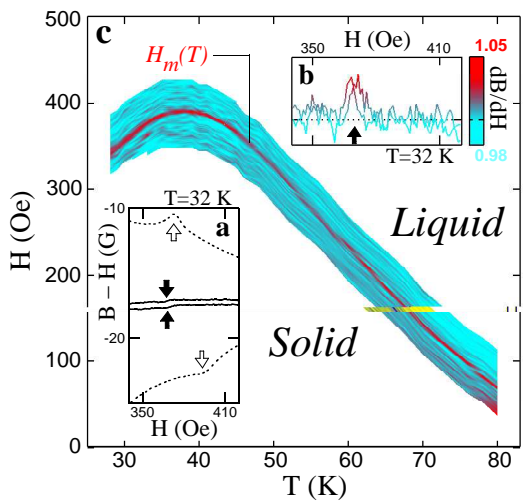}
\caption{\label{fig:hpd}(color online) (a) \mbox{32 K} field
sweeps with (solid) and without (dotted) an in-plane 350 Oe-10 Hz
`shaking' field. Reversible steps in magnetization (solid arrows)
appear instead of the hysteretic SMP (open arrows) upon shaking.
(b) The derivative of the induction with respect to applied field
$\textrm{d}B/\textrm{d}H$ at \mbox{32 K} as a color scheme. The
first-order transition appears as a paramagnetic peak (red) on top
of the $\textrm{d}B/\textrm{d}H \approx 1$ background (blue). (c)
Successive mapping of the first-order melting line
$H_\textrm{m}(T)$ measured by field sweeps.}
\end{figure}

The field sweep mapping of the FO melting line is shown in
Fig.~\ref{fig:hpd}. The collapse of the hysteretic magnetization
into a reversible behavior upon `shaking' is demonstrated in
Fig.~\ref{fig:hpd}a \short{taken at} \mbox{32 K}. \short{A
reversible magnetization} step appears instead of the SMP. To
better resolve the step we plot in Fig.~\ref{fig:hpd}b the
derivative of the measured induction with respect to applied field
$\textrm{d}B/\textrm{d}H$. The FO transition thus appears
reversibly as a $\delta$-like peak (red) on top of the
$\textrm{d}B/\textrm{d}H \approx 1$ (blue) background.
\short{Figure \ref{fig:hpd}c shows a color scheme of the
derivatives $\textrm{d}B/\textrm{d}H$ measured by field sweeps
within the temperature range 28-\mbox{80 K}. The individual
melting peaks} combine to give the locus of the FO transition line
$H_\textrm{m}(T)$. Its negative slope at elevated temperatures
becomes positive below \mbox{38 K}. This non-monotonic behavior
marks the change in the character of the transition from
thermally-induced above the extremum to disorder-driven in the
inverse melting region \cite{mikitik1, giamarchi2, nelson1}.

\begin{figure} [!b]
\includegraphics[width=0.38\textwidth]{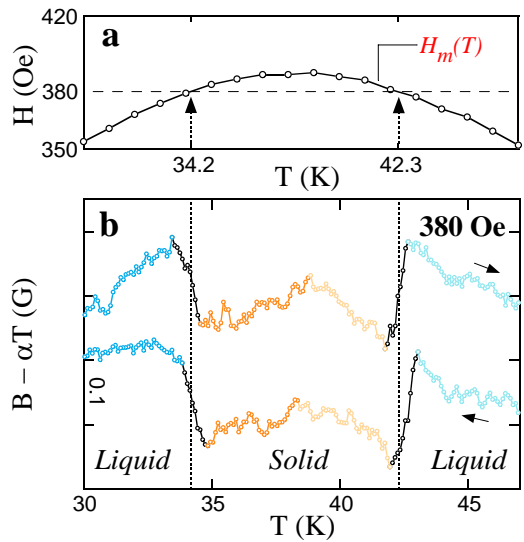}
\caption{\label{fig:380oe}(color online) (a) The FO melting line
$H_\textrm{m}(T)$ mapped via field sweeps (open circles). (b)
Local induction $B(T)$ in the presence of `shaking', measured upon
temperature sweep at \mbox{380 Oe} along the dashed line in (a). A
linear slope $\alpha T$ was subtracted for clarity. The two
equally narrow FO magnetization steps (black segments) show a
liquid-solid-liquid sequence. The temperatures at which the phase
transitions occur coincide with those, extracted from field sweeps
(doted lines). The color code reflects different phases in
\mbox{Fig.~\ref{fig:tpd}}.}
\end{figure}

In addition to the field sweeps, our experimental setup also
enables temperature sweeps in the presence of vortex `shaking'. It
thus allows to measure directly the temperature dependence of the
equilibrium magnetization at a constant applied c-axis field.
\short{At} fields slightly lower than \mbox{390 Oe}
\short{(e.g.~at \mbox{380 Oe} along the dashed line in
\mbox{Fig.~\ref{fig:380oe}a})} these sweeps should cross the
melting line twice. Remarkably, the measured local induction in
\mbox{Fig.~\ref{fig:380oe}b} indeed shows two very clear and
opposite equilibrium magnetization steps on both descending and
ascending sweeps. Note that for clarity we have subtracted from
the data a linear slope $\alpha T$. It is contributed both by the
slight temperature dependence of the Hall coefficients of the
sensors and by the linear term of the magnitude of the diamagnetic
equilibrium magnetization, which monotonically decreases with
temperature. This is the first observation of two FO transitions
obtained in a single temperature sweep. Moreover, the two steps
are equally sharp and with comparable heights of about 0.15 G.
This demonstrates that the thermally- and disorder-driven
processes, responsible for the melting and the inverse melting
respectively, are equivalent mechanisms leading to a FO
destruction of the quasi-ordered vortex solid.

The temperatures at which the magnetization steps appear along the
temperature sweep of \mbox{Fig.~\ref{fig:380oe}b} are in complete
agreement with the melting behavior deduced from field sweeps
(dotted lines). Therefore, the \short{transition line} in
\mbox{Fig.~\ref{fig:380oe}a} \short{is} independent of the
specific $H-T$ path along which \short{it is} approached - a
mandatory \short{\emph{equilibrium}} property. The remaining small
hysteresis of 0.1 to 0.2 G between downward and upward sweeps
apparently results from surface barrier effects, while the
vortices in the bulk are well equilibrated by the `shaking'. The
finite widths (about \mbox{0.7 K}) that the melting steps attain
are mainly due to a spatial and a temporal averaging mechanisms.
The first is introduced by the sensor's finite active area that
averages over the propagating melting front \cite{soibel1}, which
results from the spatially inhomogeneous equilibrium magnetization
profile \cite{zeldov1}. The temporal one is a by-product of the
`shaking' technique. The in-plane field component is known to
slightly reduce the melting temperature \cite{koshelev1, ooi1,
schmidt1}. Consequently, our time averaged measurement in the
presence of the \textsl{ac} in-plane `shaking' field results in an
additional broadening due to the instantaneous periodic shift of
the effective local melting temperature.

\begin{figure} [!b]
\includegraphics[width=0.38\textwidth]{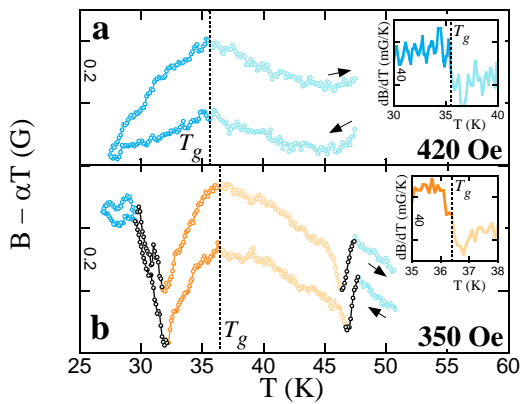}
\caption{\label{fig:350420oe}(color online) Local induction
$B(T)$, measured along temperature sweeps while `shaking'. A
reversible sharp break in the slope (at the dotted lines) appears
both above (a - \mbox{420 Oe}) and below (b - \mbox{350 Oe}) the
melting line $H_\textrm{m}(T)$. The insets show a corresponding
step in the derivative d$B/$d$T$, signifying a thermodynamic
second-order phase transition.}
\end{figure}

We thus turn to report the detection of a novel phase transition,
whose signature is a distinct break in the slope of the
magnetization $M(T)$. It is visible around \mbox{37 K} in
Fig.~\ref{fig:380oe}b at \mbox{380 Oe}, and becomes much more
pronounced at fields further away from the extremum of the FO
melting line, \short{as} depicted by Fig.~\ref{fig:350420oe}. The
\mbox{420 Oe} \short{temperature} sweep (Fig.~\ref{fig:350420oe}a)
does not intersect with the FO line, hence no steps appear in the
local induction. Nevertheless, a sharp break in slope is clearly
resolved along both descending and ascending temperature sweeps at
$T_\textrm{g}$. A \short{sharp} reversible break in the induction
slope appears also in the \mbox{350 Oe} temperature sweep of
Fig.~\ref{fig:350420oe}b (dotted line) in between the two melting
steps, hence within the solid phase. This non-analytic behavior is
emphasized by the sharp step in the derivative d$B/$d$T$ shown in
the insets. \short{These kinks were found also in other samples
and at various Hall-sensor locations, and did not depend on the
sweeping rate}. We thus conclude, that this break in slope of the
\emph{equilibrium} \short{magnetization $M(T)$} indicates a
thermodynamic \short{SO} phase transition.

\begin{figure} [!b]
\includegraphics[width=0.38\textwidth]{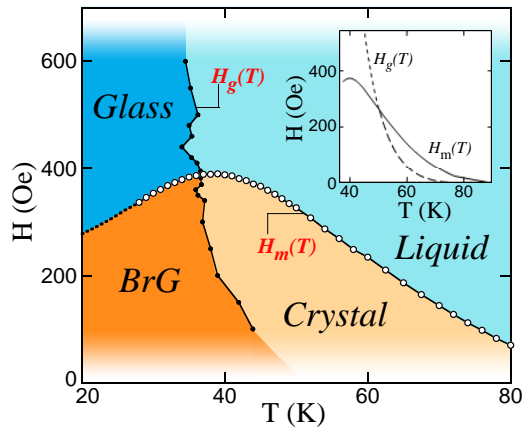}
\caption{\label{fig:tpd}(color online) The thermodynamic phase
diagram of BSCCO accommodates \emph{four} distinct phases,
separated by a first-order melting line $H_\textrm{m}(T)$ (open
circles), which is intersected by the second-order glass line
$H_\textrm{g}(T)$ (solid dots). The inset plots an equivalent
phase diagram, calculated based on Ref.~31, consisting of a
second-order replica symmetry breaking lines $H_\textrm{g}(T)$
both above (dotted) and below (dashed) the first-order transition
$H_\textrm{m}(T)$ (solid).}
\end{figure}

Mapping of both the first-order $H_\textrm{m}(T)$ and second-order
$H_\textrm{g}(T)$ transition lines onto the equilibrium $H-T$
phase diagram is given in Fig.~\ref{fig:tpd}. The SO line (solid
dots) intersects the melting curve (open circles), and shows weak
field dependence throughout the mapped region (and therefore
cannot be readily observed by field sweeps). The resulting phase
diagram consists of \emph{four} distinct thermodynamic phases.

The high-field part of the novel SO line can be naturally
identified with the long sought glass transition line. It asserts
that the low-temperature glass phase is indeed thermodynamically
distinct from the high-temperature liquid one. Several
experimental studies have observed bulk irreversibility features
in BSCCO, which appeared above $H_\textrm{m}(T)$ at about
\mbox{$35$ K} \cite{fuchs1, tamegai1}. However, all these studies
probed dynamic or non-equilibrium vortex properties. The glass
line of Fig.~\ref{fig:tpd} is the first experimental evidence of
such a thermodynamic transition in BSCCO.

Yet, the most intriguing result in Fig.~\ref{fig:tpd} is the
detection of the SO line within the vortex solid region (orange).
This implies that two distinct thermodynamic phases are present in
the low-field region below $H_\textrm{m}(T)$, contrary to the
common belief that a single BrG phase prevails throughout this
part of the phase diagram. A number of previous studies indicated
a depinning line of similar topology within the BrG below
$H_\textrm{m}(T)$ \cite{fuchs1}. However, all these measurements
probed only the non-equilibrium properties, which were consistent
with the existing theoretical dynamic predictions and simulations
of depinning \cite{nelson1, sugano1}. In contrast, the present
finding of a thermodynamic line requires a more fundamental
reconsideration.

It is interesting to note that in YBa$_2$Cu$_3$O$_7$ crystals a
thermodynamic signature of a SO transition within the liquid phase
has been reported \cite{bouquet1}. There, however, the SO line
emanates from the upper critical point of the FO line, directly
extending it to higher fields. This topology is consistent with
several dynamic measurements in YBa$_2$Cu$_3$O$_7$ \cite{safar1,
kwok1}, although alternative topologies have been also suggested
\cite{shibata1, taylor1}. In contrast, our thermodynamic data of
BSCCO show that the SO and the FO transitions are two independent
lines that intersect each other nearly at a right angle.

Several theoretical studies have shown that under the elastic
medium approximation quasi long-range order of the vortex lattice
is still retained in the presence of quenched disorder, giving
rise to the BrG phase \cite{giamarchi1, nattermann1}. This phase
was found to be stable at all temperatures (as long as topological
excitations are excluded) in systems of dimensionality greater
than two and lower than four. This is probably the reason why a
non-topological thermodynamic phase transition of the BrG phase
was hardly ever considered in 3D models. An exception is a
Josephson-glass line that was suggested to exist within the BrG
region \cite{horovitz1}. Still, the general belief is that the BrG
phase is robust until dislocations proliferate, which gives rise
to the FO phase transition \cite{giamarchi2, nonomura1,
giamarchi1}. In 2D systems, however, the BrG models did find a
possible finite-temperature depinning transition, above which
disorder is no longer relevant. Therefore, the observed SO
transition could be accounted for within these models only in the
extreme case of vanishing coupling between the superconducting
layers.

In contrast, in a recent theoretical work \cite{li1} the free
energy of the vortex matter in the presence of quenched disorder
was explicitly calculated under the lowest-Landau-level
approximation in a 3D model. It was found that two transitions are
present: a FO melting, at which the quasi long-range order is
destroyed, and a SO glass transition, below which the replica
symmetry is spontaneously broken. The inset of Fig.~\ref{fig:tpd}
shows the phase diagram, calculated using a similar effective 2D
model with parameters optimized for BSCCO \cite{rosenstein1}. The
calculations reproduce the measured features very well. Amongst
them are the melting and inverse-melting behavior, the
discontinuity of the magnetization slope $\textrm{d}M/\textrm{d}T$
at the glass transition, and the $H_\textrm{g}(T)$ line itself,
which resides at slightly higher temperatures as compared to
experiment. In addition, the calculations show that the portions
of the $H_\textrm{g}(T)$ line, lying above and below the FO
transition $H_\textrm{m}(T)$ (dotted and dashed lines,
respectively), are slightly shifted from each other. Therefore,
they do not cross the melting line at a single point, but rather
form two closely located tricritical points along it. This minute
shift can hardly be seen in the inset of Fig.~\ref{fig:tpd}, and
is below our current experimental resolution.

Within this model the high-field glass phase and the low-field BrG
are strongly pinned and replica symmetry broken, whereas the two
high-temperature phases are replica symmetric and thus reversible.
This conclusion of reversibility is consistent also with the
existing dynamic measurements \cite{fuchs1}. It is therefore
tempting to speculate that the phase above $H_\textrm{g}(T)$  and
below $H_\textrm{m}(T)$ should acquire a true crystalline order.
However, since Rosenstein and Li did not calculate the structure
factor and our measurements do not probe this quantity, the
proposed vortex crystal phase certainly calls for further
experimental and theoretical investigations.

In summary, we present thermodynamic evidence of a possibly
second-order glass transition line that splits the quasi-ordered
vortex solid into two distinct phases. By comparing the results
with existing dynamic measurements and a new theoretical study, we
suggest that the two phases are BrG and a vortex crystal. The
glass line crosses the first-order melting line near its extremum
and extends to higher fields, giving rise to two thermodynamically
distinct disordered phases - a glass and a liquid.

We thank D.~Li, V.M.~Vinokur, and B.~Horovitz for stimulating
discussions. This work was supported by the Israel Science
Foundation Center of Excellence, by the German-Israeli Foundation
G.I.F., by the Minerva Foundation, Germany, and by Grant-in-aid
for Scientific Research from the Ministry of Education, Culture,
Sports, Science, and Technology, Japan. BR acknowledges the
support of the Albert Einstein Minerva Center for Theoretical
Physics and EZ the US-Israel Binational Science Foundation (BSF).

\end{document}